\documentclass[iop]{emulateapj}
\usepackage{graphicx}
\usepackage{natbib}

\def\deg{$^{\circ}$}

\def\kmskpc{km s$^{-1}$ kpc$^{-1}$}

\shorttitle{Orbital Support of Inner Bars}
\shortauthors{Witold Maciejewski and Emma E. Small}

\submitted{}

\begin{document}

\title{Orbital Support of Fast and Slow Inner Bars in Double Barred Galaxies}

\author{Witold Maciejewski and Emma E. Small}
\affil{Astrophysics Research Institute, Liverpool John Moores University, Twelve
Quays House, Egerton Wharf, Birkenhead, CH41 1LD}
\email{wxm@astro.livjm.ac.uk}

\begin{abstract}
We analyze how the orbital support of the inner bar in a double-barred 
galaxy (nested bars) depends on the angular velocity (i.e. pattern speed) of 
this bar. We study orbits in seven models of double bars using the method of 
invariant loops. The range of pattern speed is covered exhaustively.
We find that not all pattern speeds are allowed when the inner bar 
rotates in the same direction as the outer bar. Below a certain minimum pattern
speed orbital support for the inner bar abruptly disappears, while at
high values of this speed the orbits indicate an increasingly round bar that
looks more like a twist
in the nuclear isophotes than a dynamically independent component.
For values between these two extremes, orbits supporting the inner bar extend
further out as the bar's pattern speed decreases, their corresponding loops
become more eccentric, pulsate more, and their 
rotation becomes increasingly non-uniform, as they speed up and slow down 
in their motion. Lower pattern speeds also lead to a less 
coherent bar, as the pulsation and acceleration increasingly varies 
among the loops supporting the inner bar. The morphologies of fast and
slow inner bars expected from the orbital structure studied
here are recently recovered observationally
by decomposition of double barred galaxies. Our findings 
allow us to link the observed morphology to the dynamics of the inner bar.
\end{abstract}

  
\section{Introduction}
Double barred galaxies are barred spiral galaxies that contain an additional
small bar nested within the main bar. First observed by de Vaucouleurs (1975),
they may constitute $\sim$ 30\% of barred galaxies, as implied by photometric 
surveys (Erwin \& Sparke 2002, Laine et al. 2002). However, 
cross-correlation of these surveys suggests a lower percentage (Moiseev 
2010), which is consistent with a kinematic survey of double 
barred candidates using integral field spectroscopy (Moiseev et al. 2004).
The frequency at which double bars are observed indicates that 
they are either recurrent or long-lived phenomena. Since inner bars occur 
often in early-type galaxies with little or no gas to drive evolution, 
they are most likely long-lived. This is consistent with the fact that 
inner bars are detected in both optical and IR observations, indicating that 
they are stellar structures. Observations of the apparent random orientations of the two bars
suggest that the bars are rotating independently (Buta \& Crocker 1993; Friedli \& Martinet 1993).
This was later 
confirmed for the galaxy NGC 2950, whose inner and outer bar do not rotate at the same pattern 
speed (Corsini et al. 2003).

It is difficult to understand how two independently rotating nested bars can be maintained by 
regular motions of stars to create a long-lived stable system. In particular, the resonances 
of one bar will interfere with the orbital support of the other, which is likely to produce
chaotic motions and thus constrain possible parameters of stable self-consistent 
double bars. However, when studying the orbital 
response to an assumed potential of two independently rotating bars Maciejewski \& Sparke
(2000, hereafter MS00) found stable 
orbits that can support each bar in its rotation. 
Maciejewski \& Athanassoula (2007, hereafter MA07)
showed that double bars are sustained by families of stable double 
frequency orbits, i.e. orbits that oscillate only with the driving frequencies of the 
two bars. Support for the two bars is provided by trajectories trapped around these double
frequency orbits. 

Maciejewski \& Athanassoula (2008, hereafter MA08) studied the trapping of trajectories around
regular orbits in 23 models of double bars, by varying the parameters characterising both
bars. They varied the lengths, masses and eccentricities of the bars, but they found
that out of the parameters of the inner bar, its pattern speed affected the trapping most.
In this paper we investigate how the appearance of double frequency orbits supporting the 
inner bar changes with its pattern speed. Thus we perform the orbital structure studies, 
following the method developed by Contopoulos \& Papayannopoulos (1980), 
Athanassoula et al. (1983), Teuben \& Sanders (1985) and others. 
These studies do not aim to construct self-consistent models, 
but rather explore changes in the system that occur when its main parameters vary.

We study orbits in Models 01 -- 05 from MA08, where the pattern speed
of the inner bar was varied between 80 \kmskpc\ and 120 \kmskpc. We also construct new 
models 02E and 05E here, which extend this range down to 70 \kmskpc\ and up to 140 \kmskpc. 
Thus all models considered in this paper are identical except for the pattern speed of the
inner bar. They are the same as Model 2 from MS00, where
the semi-major axis of the outer bar is 6 kpc and the semi-major axis of the inner bar
is 1.2 kpc. The axial ratios of the outer and inner bar are 2.5 and 2.0,
respectively. The mass of the inner bar is 15\% of the mass of the outer bar. The 
pattern speed of the outer bar is 36 \kmskpc\ and both bars rotate in the direction of
the stellar rotation in the disk. 

In Section 2, we briefly review the method that uses invariant loops in the search for regular
orbits in double bars, and we describe how we derive the position angle and eccentricity
of the loops as well as the extent to which they support the inner bar. In Section 3, we 
analyze the orbital structure in each of the seven models, and we find what
trends in the parameters and behaviour of the inner bar, as a function of its pattern 
speed, its orbital structure indicates. We discuss the validity of orbital structure studies
in Section 4, where we also attempt to obtain information on the dynamics of the inner bar 
from its observed characteristics. We summarize our findings in Section 5.

\section{Method}
Like stable closed periodic orbits in a single bar, stable double frequency orbits in double 
bars provide support for both bars by trapping trajectories that oscillate around them.
The difficulty in studying double frequency orbits within two independently rotating bars
is that such a potential oscillates in time, and these orbits do not close. However, double 
frequency orbits can be visualised using the concept of a loop, invented by Maciejewski \&
Sparke (1997). A coherent tutorial on the method based on invariant loops can be found in
MA07 and Maciejewski (2010); here we outline its basic concepts. Any orbit 
(which consists of particle positions over a continuum of moments in time) can be represented 
for any instantaneous shape of the oscillating potential by points on this orbit plotted at 
the moments when the potential repetitively takes that given shape (a discreet set of moments
in time).  
In particular, double frequency orbits are represented by closed curves, called loops (MA07). 
The loops change shape as the bars rotate through each other, responding to the oscillating
potential, but they regain their former shape when the bars return to their former
relative orientation. The shape of the loop does not depend of the reference frame in which 
the loop is drawn, hence the loops are sometimes called the invariant loops.

\begin{figure*}
\centering 
\includegraphics[scale=0.85]{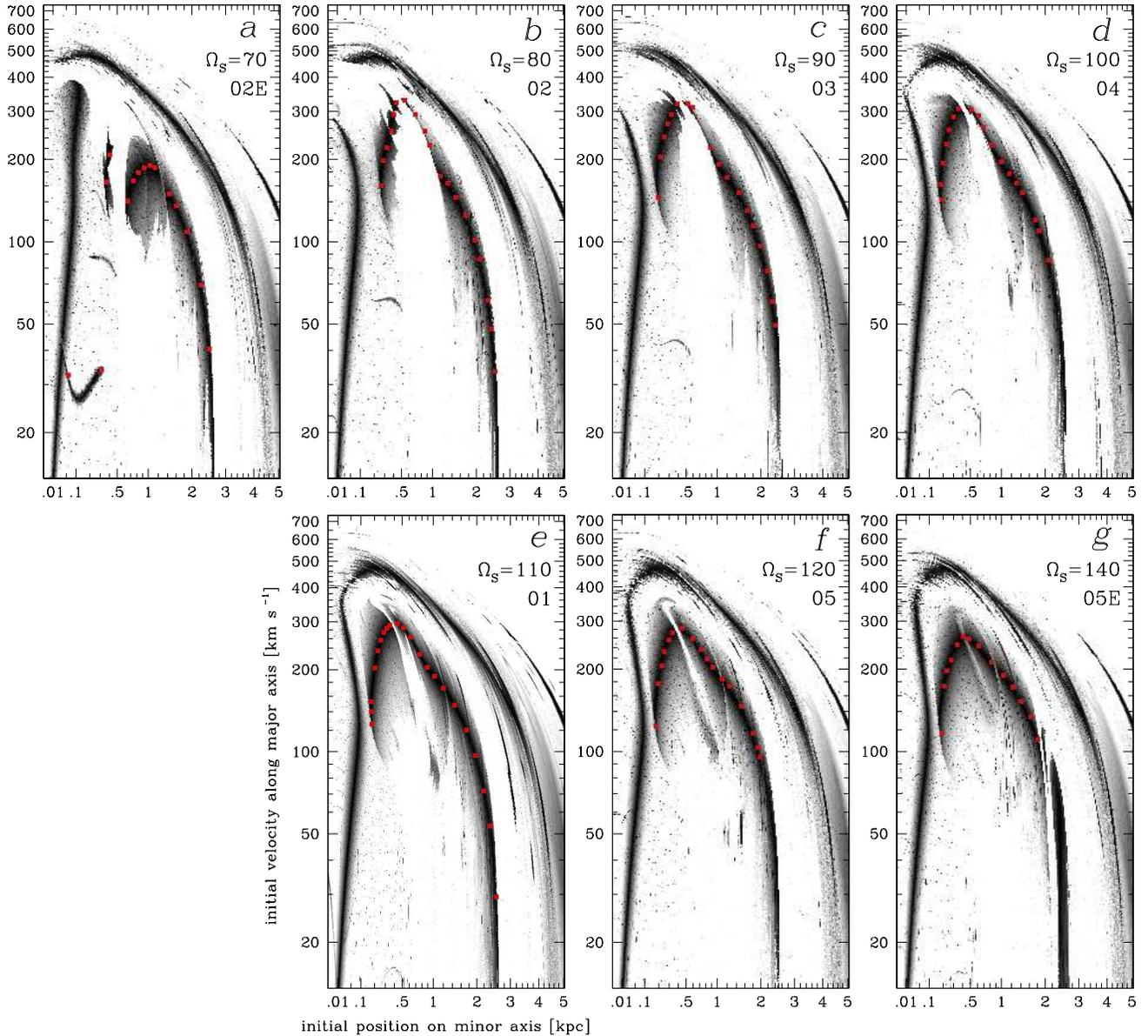}
\vspace{0.05cm}
\caption{Diagrams of ring-width for all the models (02E, 02, 03, 04, 01, 
05 and 05E -- ordered by increasing pattern speed of the inner bar) 
as a function of initial position along the minor axis of the aligned bars 
(horizontal axis) and the initial velocity along the major axis (vertical axis).
Darker regions represent smaller ring-widths. The red points correspond to the initial 
conditions of representative $x_2$ loops plotted in Figure 2, and analysed in Figures 3 and 4. 
In the top-right corner of each frame, we list the pattern speed of the inner bar 
($\Omega_S$, in \kmskpc), and the model number.}
\vspace{0.1cm}
\end{figure*}

\begin{figure*}
\centering
\vspace{-0.5cm}
\includegraphics[scale=0.92]{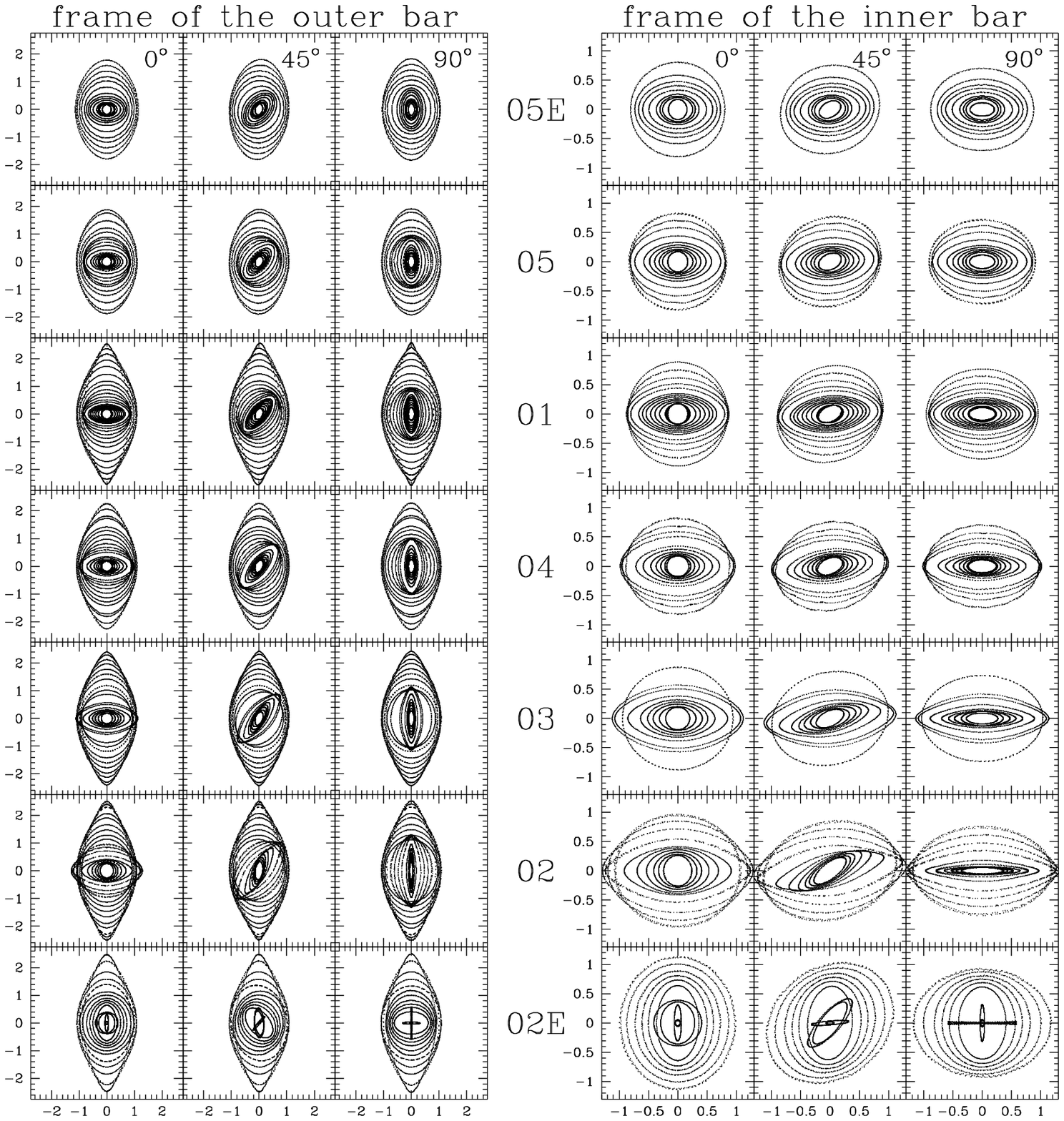}
\vspace{-1.2cm}
\caption{Representative $x_2$ loops in all models, whose initial conditions are marked by 
the red dots in Figure 1. The loops in the three left-hand columns are plotted in the frame 
of the outer bar, in which the outer bar remains horizontal and the inner bar rotates
anticlockwise. In the three right-hand columns, the inner loops out to the most 
circular loop are plotted in the frame of the inner bar, where the inner bar remains 
horizontal. For each reference frame the 1st, 2nd and 3rd panels in each row
show loops when the angle between the bars in the imposed potential is 0\deg, 
45\deg\ and 90\deg, respectively. The rows show, from top to bottom,
models 05E, 05, 01, 04, 03, 02 and 02E. The units on the axes are kpc.}
\end{figure*}

Trajectories trapped around a double frequency orbit 
are represented by rings that enclose the loop 
which represents that orbit. Thinner rings indicate smaller amplitudes of oscillations 
and a more
tightly bound trajectory (MA07). MA08 constructed ring-width diagrams, which display the 
thickness of the rings as
a function of the trajectory's initial position along the minor axis of the aligned bars and 
initial velocity along the major axis. Trajectories in the plane of the galactic disk 
can be determined
from these two starting conditions if it is assumed that they begin on the minor axis
with no 
radial velocity component. MA08 showed that this 
subset of initial conditions reproduces all major families of regular orbits in double bars.

\begin{figure*}[t]
\centering
\vspace{-0.5cm}
\includegraphics[scale=0.9]{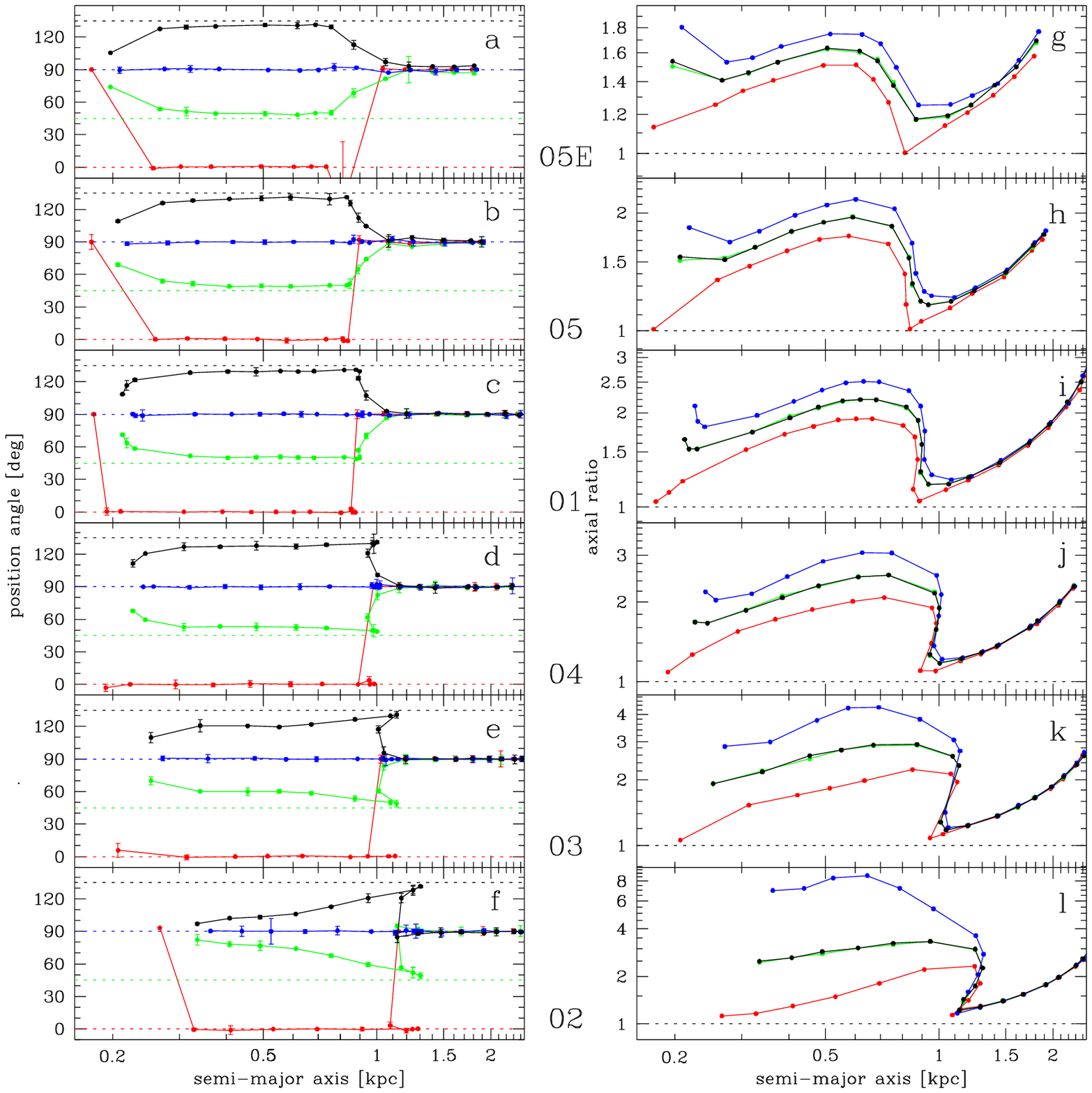}
\vspace{-1.2cm}
\caption{For all representative $x_2$ loops in models 05E, 05, 01, 04, 03 and 02 (top to bottom),
the position angle (PA, in degrees) in the frame corotating with the outer bar in which that bar 
remains horizontal (left-hand panels) and the axial ratio (right-hand panels) are plotted 
against the length of the semi-major axis (in kpc). Plots are for the relative angle between 
the bars in the imposed potential equal 0\deg\ (red), 45\deg\ (green), 90\deg (blue) and 
135\deg\ (black). Lines that connect points representing loops reflect the sequence along the 
arch in Figure 1.}
\end{figure*}

Figure 1 displays the ring-width diagrams for all the models of double barred 
galaxies analysed in this paper. These are models 01 -- 05 from MA08, 
whose ring-width diagrams have been presented there, and new models 02E 
and 05E, whose ring-width diagrams are constructed here using the same algorithm.
The darker regions represent smaller ring-widths, hence initial conditions
from the darkest spine of the two arches generate a close approximation to 
double frequency orbits. The outer arch corresponds to orbits that support
the outer bar and originate from the $x_1$ orbits in a single bar. The inner arch 
corresponds to orbits that originate from the $x_2$ orbits of a single (outer) 
bar. Loops that represent these orbits vary between 
those which are perpendicular to the outer bar, and those which support the inner bar, with 
continuous transition between these two extremes. Because of this continuity, 
these orbits belong to a single family, which we call the $x_2$ family. Note
that loops which support the inner bar (sometimes vaguely called 'the inner 
bar's $x_1$ orbits') are called in our notation the $x_2$ loops, which reflects their
origin.

In order to recover the orbits supporting the inner bar, we sample the 
inner arch uniformly along the horizontal axis of the diagram (which is proportional to 
the square root of the initial position coordinate), choosing initial conditions that return the
smallest ring-widths. These initial conditions, marked in Figure 1 by red dots, only approximate 
the double frequency orbits,
with accuracy proportional to the resolution of the ring-width diagram. This accuracy is
sufficient for the purpose of this paper. Loops representing such approximate double
frequency orbits are constructed by following a particle on this orbit for 399
successive realignments of the bars. Thus each loop is made up of 400 points. 
For each of the models, loops whose initial conditions are marked by red dots in Figure 1
are plotted in Figure 2 for three different relative angles between the two bars in the imposed 
potential, and diagrams characterizing properties of these loops are presented in Figures 3 and 4.

\subsection{Determination of the position angle (PA) of the loops}
We define the position angle (PA) of the semi-major axis of each loop as the PA
at the maximal radial coordinate on this loop. This method works well
when loops are close to elliptical (m=2 component dominating) but will return incorrect 
semi-major axes
for loops that are more rectangular (m=4 component dominating) or that are not symmetrical
at all. However all the loops analysed here appear to be very ellipse-like. 
To test the validity of this analysis we reflect the points of the inner most and
most circular loop of model 01 through its derived semi-major axis. 
The innermost loop was chosen in order to ensure it was symmetrical since it appears to be
affected by both bars and is aligned with neither. For both orbits there was a 
small deviation in the reflected points compared to the original, 
but it was of the order of the width of the ring that best approximates its loop.
This confirms the validity of the method. The method is subject to error arising from 
insufficient sampling or scattering of points near the real semi-major axis. 
We estimate the error as the difference between the two estimates of the PA
of the semi-minor axis: one being the PA at the minimum radial 
coordinate, and the other the PA of the semi-major axis subtracted by 90\deg.
The error is expected to be
greater for more circular loops, where the estimation of the semi-major axis is more affected
by scatter. The PAs for all the loops in each model, together with estimated error
are plotted in the left panels of Figure 3 for four relative positions of the bars in the
imposed potential: 0\deg, 45\deg, 90\deg\ and 135\deg. 

\subsection{Determination of the eccentricity of the loops}
The eccentricity of a loop is quantified by its axial ratio: the ratio between the lengths
of its semi-major and semi-minor axes. The semi-minor axis here is taken as the 
radial coordinate of the point located 90\deg\ away from the previously 
determined semi-major axis. 
Unlike the PA, the actual lengths of the semi-major and semi-minor axes are
not appreciably affected by the scatter and the error in axial ratio is small. 
The axial ratios for all loops in each model are plotted in the right panels of Figure 3 for
four different relative positions of the bars in the imposed potential: 0\deg, 45\deg, 
90\deg\ and 135\deg. The axial ratios at 45\deg\ and 135\deg\ are the same, which 
reflects the symmetry of the problem upon time reversal. The amplitude at which the 
loops pulsate can be estimated from the variation in eccentricity of the loops as the
bars rotate. From Figure 2, we see that the eccentricity of the loops is smallest when
the bars are aligned and largest when the bars are perpendicular. The ratios of eccentricity of 
loops, when the bars in the imposed potential are separated by 45\deg and 90\deg, to 
their eccentricity, when the bars are aligned, are displayed for all models in Figure 4.

\begin{deluxetable*}{lccccccc}
\centering
\tabletypesize{\scriptsize}
\tablecaption{Characteristics of models of double bars}
\tablehead{
\colhead{Model name} & \colhead{$\Omega_S$ [\kmskpc]} & \colhead{$r_{CR}$ [kpc]} & \colhead{$a_0$ [kpc]} & \colhead{$a_0/r_{CR}$} & \colhead{$e_0^{\rm max}$} & \colhead{$e_{90}/e_0$}      & \colhead{$\delta PA$ [degrees]}
}
\startdata
02   &  80        & 2.92     & 1.284 & 0.440        & 2.32            & 3.68 (1.46--6.46) & 18.55 (3.63--33.12)\\
03   &  90        & 2.63     & 1.117 & 0.425        & 2.23            & 1.88 (1.39--2.33) & 10.82 (3.92--15.42)\\
04   & 100        & 2.39     & 0.984 & 0.412        & 2.08            & 1.40 (1.27--1.53) &  6.44 (3.72--8.34)\\
01   & 110        & 2.19     & 0.877 & 0.401        & 1.91            & 1.29 (1.23--1.32) &  5.26 (3.88--6.70)\\
05   & 120        & 2.01     & 0.812 & 0.403        & 1.75            & 1.23 (1.20--1.25) &  5.47 (3.66--9.09)\\
05E  & 140        & 1.74     & 0.735 & 0.422        & 1.51            & 1.18 (1.15--1.22) &  5.20 (3.21--8.73)\\
\enddata

\tablecomments{
Col.2: pattern speed of the inner bar in the assumed potential (i.e. the input parameter of the model).
Col.3: corotation radius of the inner bar.
Col.4: the length of the semi-major axis of the inner bar as determined from the extent of the 
loops that support it, measured at the moment when the two bars are parallel.
Col.5: the ratio of the length of the inner bar to its corotation radius.
Col.6: the largest axial ratio of the loops supporting the inner bar at the moment when the two 
bars are parallel.
Col.7: average ratio of the eccentricity of the loops supporting the inner bar at the moment when 
the two bars are perpendicular to the eccentricity when the two bars are parallel (in parenthesis 
the range of this ratio for all the loops that support the inner bar).
Col.8: the average offset of the PA of the loops supporting the inner bar from the 
PA of the inner bar in the assumed potential being 45\deg\ (in parenthesis the range 
of this offset for all the loops that support the inner bar).}
\end{deluxetable*}

\subsection{Determination of the length of the inner bar}
Although by assuming the potential in which we calculate the orbits 
we thereby set the lengths of the two bars, we have to analyse the orbital 
response in order to estimate to what extent the orbits support
the bars. Loops that represent orbits supporting a given bar should
follow that bar as the bars rotate through each other. For most models 
in Figure 2, the inner $x_2$ loops generally do seem to follow the inner bar,
while the outer $x_2$ loops remain perpendicular to the outer bar. There 
is a smooth transition between the inner and outer $x_2$ loops, where
loops are almost circular. This is confirmed by the PA plots in the
left-hand panels of Figure 3, where the PAs of the outer loops remain
consistently at 90\deg\ regardless of the relative position of the
bars, while the PAs of loops with semi-major axes roughly between
0.3 kpc and 0.8 kpc or more (depending on the model) are also consistent
with each other, but their value changes in accordance with the rotation
of the inner bar in this frame. Departures from this simple picture,
caused by different behaviour of up to the three most inner loops in all models and by incoherent
PAs of inner loops in models with lower inner bar's pattern speeds, will 
be analyzed in Section 3.1.

In Figure 1, the initial conditions for the $x_2$ loops come from the darkest
spine of the inner arch, and therefore the loops can be ordered into a sequence
along this arch. This sequence is reflected in Figure 3 by lines which connect 
points marking individual loops. We 
determine the last loop supporting the inner bar as the last of the loops
that maintain a consistent PA, which varies in accordance with the PA of the
inner bar in the imposed potential. Then among the loops supporting the inner
bar we find the one whose major axis is longest, and the length of the 
bar is defined as the length of this major axis. Note that the loop with 
the longest major axis does not have to be the last one in the sequence 
of loops supporting the bar. As can be seen in Figure 3, in models with lower 
angular velocity of the inner bar (lower panels), the semi-major axis of the 
loops which support that bar initially increases along the sequence defined 
by the arch in Figure 1, but then reaches a maximum and decreases, so that the 
last of the loops supporting the bar is not the loop of the longest semi-major 
axis. 

Since the loops presented here are only a representative sample of the $x_2$ orbital 
family, the definition formulated above underestimates the length of the inner 
bar. The upper limit for this length can be estimated in two ways, depending on 
whether the loop with the longest semi-major axis is the last one in the sequence 
of loops supporting the bar. If it is, then the bar is shorter than the semi-major 
axis of the next loop in the sequence past the last one supporting the bar. 
Otherwise, the upper limit for the bar length is calculated by adding to the 
longest among the semi-major axes of loops that support the bar an average 
spacing in the semi-major axis between the loops that precede and follow it.

\begin{figure}
\includegraphics[width=1.0\linewidth]{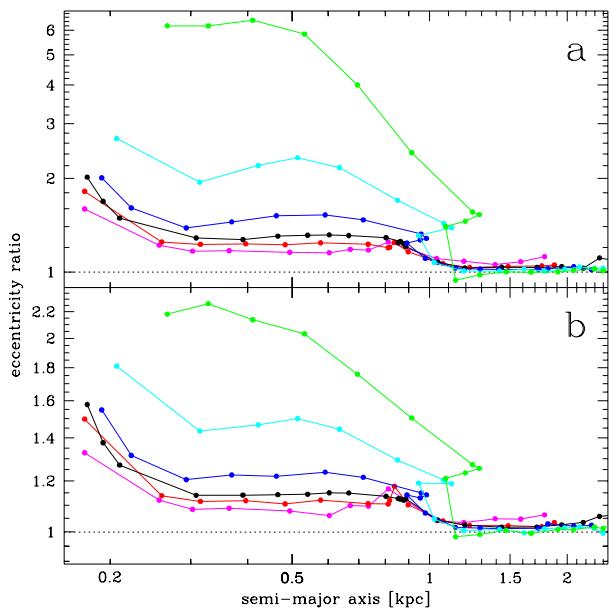}
\caption{The ratio of eccentricity of the loops, when the angle between the 
bars in the imposed potential is 90\deg\ (top panel) and 45\deg\ (bottom panel),
to their eccentricity when the bars are aligned, plotted against their semi-major 
axes measured when the bars are aligned. Each representative loop in models 02-05E
is marked by a point, with model 05E being plotted in pink, 05 -- in red, 
01 -- in black, 04 -- in blue, 03 -- in cyan and 02 -- in green. Lines that 
connect points representing loops reflect the sequence along the arch in Figure 1.}
\end{figure}

\section{Results}
There is a gradual change in the global orbital structure of the calculated models,
as the angular velocity of the inner bar is varied. These global properties are
quantified and listed in Table 1 for all models in which the inner bar has orbital 
support. The pattern speed of the inner bar, listed in the second column, is the input parameter 
of the model. The corotation radius of the inner bar, listed in the third column, is calculated 
in the axisymmetric approximation, and therefore depends only on the radial mass distribution 
in the model. The parameters listed in columns 4--8 are extracted from the orbital structure 
calculated for every model. The variation of these parameters with the pattern speed of the inner
bar is studied in the second half of this section, while in the first half we remark on the
details of finding regular orbits in each model, and on extracting from them the numerical
parameters for Table 1.

\subsection{Notes on individual models}
\subsubsection{Model 01}
We start with model 01 because it is the reference model in MA08 and MS00. The two bars 
in model 01 are in resonant coupling with the corotation of the inner bar overlapping 
the Inner Lindblad Resonance (ILR) of the outer bar. MA08 calculated the ring-width 
diagram for this model with the resolution twice higher than that for the other models 
(i.e. twice denser sampling of the initial conditions on both axes), therefore the parameters 
of the loops can be determined here more accurately, giving closer approximations to double 
frequency orbits and hence smoother loops in Figure 2. On the example of model 01 we show here 
how the parameters of the loops are extracted.

The last loop
supporting the inner bar in model 01 is the 12th loop. According to the definition adopted in 
Section 2.3, the length of the inner bar is estimated by the longest major axis  
among the loops that support it. In model 01 it is also the 12th loop.
It is evident from the PA diagram in Figure 3c that almost all $x_2$
loops, up to the radial extent determined above, are well aligned with each 
other for each of the four relative positions of the bars in the
imposed potential. This indicates that the bar maintains its shape coherently 
throughout rotation. However, the three 
inner-most loops are not aligned with the loops supporting the inner bar, when 
the angle between the two bars in the imposed potential is 45\deg\ and 135\deg. 
The acceleration of  these loops is more extreme and, as we see below, the 
inner-most loop is actually dominated by pulsation.
The initial conditions for these loops were taken from the edge of the 
left leg of the inner arch in the ring-width diagram (Figure 1e). They are close 
to the inner ILR of the outer bar and are most likely affected by this resonance. 
Since the inner-most regions of the inner bar can only be supported when the
$x_2$ orbits extend to the galaxy centre, there may be no inner ILR in self-consistent
double barred galaxies, and the behaviour of the innermost loops in model 01
is likely to be an artifact caused by the potential assumed in this work. 
Below we exclude the inner-most loops when estimating the PA and pulsation 
of the inner bar.    

The acceleration and pulsation of the loops supporting the inner bar, previously seen 
by MS00, is clear in the right-hand panels of the 3rd row in Figure 2.
These loops are aligned with the inner bar in the imposed potential when the angles 
between the bars there are 0\deg\ and 90\deg\ but they lead when this angle is 45\deg\ 
and trail when it is 135\deg. This is confirmed by the PA diagram in Figure 3c:
when the angle between the two bars in the imposed potential is 45\deg, the mean
PA of the loops supporting the inner bar in model 01 (loops 4-12) is 50.3\deg\
with average error $\sim$1.1\deg. The pulsation of the inner bar
is evident in Figure 2 from the change in the appearance of the
loops between the time when the bars are parallel and perpendicular to each other. 
The loops are most eccentric when the bars are perpendicular. This is also shown in Figure 3i: 
the axial ratio of the loops supporting the inner bar increases
when the two bars are not aligned. These loops seem to be pulsating consistently 
since their eccentricities change at the same rate
(Figure 4), aside from the three innermost loops, which strongly pulsate.

The transition between loops supporting the inner bar and those perpendicular to the outer
bar occurs for initial conditions located in a very smooth part of the inner arch in the
ring-width diagram (Figure 1e), between initial positions along the minor axis of
0.65 and 1.2 kpc. At smaller initial positions there is a discontinuity in the inner arch.
MA08 suggested that this feature was associated with the transition region, but it occurs
between the 9th and 10th loop, which is in the middle
of the orbits supporting the inner bar, and its origin is still unclear.

\subsubsection{Model 05}
The pattern speed of the inner bar in models 05 and 05E is successively higher 
than in model 01. In model 05 it is 120 \kmskpc. In this model, 
the loops supporting the inner bar do not extend as far as they do in model 01. 
The last loop supporting the inner bar is the 8th loop, and this loop also has
the longest semi-major axis out of the loops that support the inner bar. 
As in model 01, the loops supporting the inner bar behave coherently in this model, 
which can be seen in Figure 3b. The inner-most loop is misaligned and dominated by pulsation
and is therefore excluded from supporting the inner bar. 
The mean PA of the loops that support the inner bar (loops 2-8) 
when the bars in the imposed potential are separated by 45\deg\ is 50.5\deg 
with an average error of 1.6\deg. This implies
the acceleration of the inner bar in model 05, within the errors, the same as in model 01.
The axial ratio plot in Figure 3h shows that the loops are
rounder in model 05 than in model 01. Variation in eccentricity of the loops (Figure 4, Table 1)
implies that the inner bar in model 05 pulsates less than in model 01. 

\subsubsection{Model 05E}
The pattern speed of the inner bar for this model is 140 \kmskpc. This is a new model, 
not presented in MA08, and its ring-width diagram is shown in Figure 1g.  
The discontinuity at the top of the inner arch in this diagram is reduced when compared to the 
other models. 
There is also more grey area surrounding the inner arch, which indicates a greater phase-space 
volume occupied by trapped orbits that can provide support for the inner bar.
In this model, the loops 
supporting the inner bar are completely nested within the loops that are perpendicular
to the outer bar. There is no gathering of loops at the 
end of their semi-major axes as there is in the other models,
which makes the extent of the support for the inner bar less well defined in model 05E. 
The last loop supporting the inner bar (the 8th loop) has also
the longest semi-major axis, and therefore serves as the estimate of the bar length.

As in models 01 and 05, the loops supporting the inner bar in model 05E (loops 2-8, 
the first loop is excluded because of pulsation) maintain coherence throughout 
rotation (see Figure 3a). When the angle between the bars in the imposed potential
is 45\deg, the mean PA of loops supporting the inner bar is 50.2\deg, 
with an average error of 1.5\deg. This implies that the acceleration of the inner bar
in this model is consistent with that in models 01 and 05. The loops in model 05E 
are the roundest of all models, as shown in Figure 3g. The eccentricity of the loops
(Figure 4) implies that the inner bar in model 05E
also seems to undergo less pulsation than in any of the other models. 

\subsubsection{Model 04}
In models 04 to 02E, the pattern speed for the inner bar successively decreases 
from its value in model 01. In model 04 it is 100 \kmskpc.
The extent of the support for the inner bar in this model is greater
than in model 01. The last loop supporting the inner bar in model 04 is 
the 10th loop, but among these loops the 9th loop serves as the estimate of the
bar length, because it has the longest semi-major axis.

As can be seen in Figure 3d and Figure 4, the loops supporting the inner bar 
in model 04 (loops 3-10, since as in the models above, two inner-most orbits,
dominated by pulsation, are excluded from loops supporting the inner bar)
still behave coherently throughout rotation.
The average PA of loops supporting the inner bar, when the angle between 
the bars in the imposed potential is 45\deg, is 51.4\deg
with an average error of 2.5\deg. This implies the acceleration of loops 
in model 04 slightly larger than in the models above, but by no more
than a typical measurement error. Figure 3j shows that the loops supporting 
the inner bar in model 04 are more eccentric than in model 01. The loops 
supporting the inner bar in model 04 also undergo a
lot more pulsation than in model 01 (Figure 4).

\subsubsection{Model 03}
The pattern speed of the inner bar for model 03 is 90 \kmskpc. 
There is a gap between loops 8 and 9, which corresponds 
in the ring-width diagram (Figure 1c) to the region of the inner arch, where its
dark spine is apparently missing. We could not find good approximations to the
loops in this region. This means that even if a more detailed search succeeds in
finding the loops there, regular orbits are not trapped well there, hence the 
support for the inner bar is strongly reduced. Nevertheless, the extent 
of the inner bar, in terms of support from the $x_2$ orbits,
is still greater than in the models above. 
The last loop supporting the inner bar is the 8th loop, and because of the gap
past it, where we could not find loops, it has the longest semi-major axis among 
these loops. However, the extent of the bar may
still be slightly larger if a loop supporting this bar were found past the 8th
loop.

Contrary to the models analysed so far, the loops supporting the inner bar in 
model 03 do not behave coherently throughout rotation: when the angle between the 
bars in the imposed potential is 45\deg, in Figure 2 there is a twist in loops supporting 
the inner bar, reflected by a change in the PA in Figure 3e. One can see that
the deeper inside the bar the loop is located, the more it accelerates.
An upturn in the PA for the inner-most loop appears now to be more consistent with the trend
exhibited by the other loops, but for the sake of consistency with the other models
we still exclude the inner-most loop from the loops supporting the inner bar.
In model 03, the mean PA of the loops supporting the inner bar (loops 2--8) is larger 
than in the models above (Table 1), hence the inner bar in this model undergoes more acceleration.
The loops supporting the inner bar in model 03 are more eccentric compared to 
the models above and they imply greater pulsation of the inner bar than in the models above,
but this pulsation is no longer consistent among the loops (Figure 4). 

\begin{figure}[t]
\centering
\includegraphics[width=1.1\linewidth,viewport=10 260 600 550,clip]{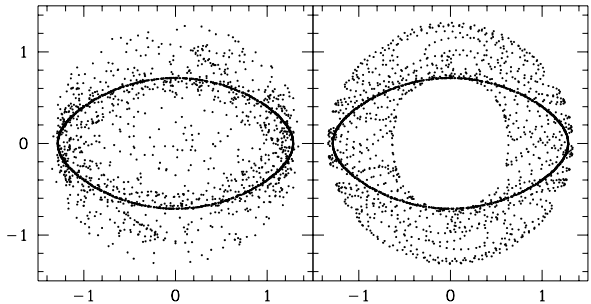}
\caption{Representations of three orbits in the potential of two independently 
rotating bars from model 02, plotted at the moments when the two bars are aligned, 
for 1200 consecutive alignments of the bars. The solid line in both panels marks 
the 8th loop in this model. The dots in the left-hand panel mark the representation
of an orbit with the same initial velocity and the initial position equal to 0.97 of that
of the orbit that is represented by the 8th loop. The dots in the right-hand panel 
mark the representation of an analogous orbit whose initial position is now multiplied
by 1.03. Units on axes are in kpc.}
\end{figure} 

\subsubsection{Model 02}
The pattern speed of the inner bar in model 02 is 80 \kmskpc. The last among the loops 
supporting the inner bar is the 9th loop, but among them the 8th one has the longest 
semi-major axis and therefore serves as the bar length estimate.
In the ring-width diagram in Figure 1b, the initial conditions for loops 8-10 in 
model 02 are taken from a ridge-like feature in the right leg of the inner arch. 
This ridge departs upwards from the course of the dark spine of 
the inner arch on the lower right. On the right of this ridge, 
there are light grey areas indicating poorly trapped trajectories, and on the 
left there is white area indicating chaotic trajectories with initial conditions 
immediately next to 
those of the loops. Finding loops on this ridge is unexpected since stable double 
frequency orbits should be well surrounded by trapped trajectories. In fact in 
model 03, for corresponding initial conditions (between loops 8 and 9 there) we were 
not able to find loops at all: trapping of trajectories was increasing when 
moving to the left in the ring-width diagram in this region, but then the trajectories were
abruptly becoming chaotic before they converged on loops. We checked the 
appearance of orbits around the ridge in model 02: Figure 5 shows the 8th loop, and points
sampled from the trajectories whose initial positions are just 3\% 
smaller and larger than that of the loop (i.e. come from adjacent pixels in the ring-width
diagram). As expected, the trajectory with initial conditions from the left 
of the ridge (Figure 5, left-hand panel) looks chaotic, while the trajectory 
whose initial conditions are to the right of the ridge (Figure 5, right-hand panel),
although regular, does not align with the loop, and hence it does not support 
the bar. Therefore the outermost loops of the inner bar (loops 7-9) in model 
02 provide essentially no support for this bar in terms of trapped trajectories. 
The resolution of the ring-width diagram is barely sufficient to locate these
loops, hence in Figure 2 they appear as a collection of noticeably scattered points.

The twisting of the loops in the inner bar is even more dramatic in model 02 than in model 03.
In Figure 2, these loops form a spiral rather than a bar,
when the angle between the bars in the imposed potential is 45\deg. As in model 03, 
the inner-most orbit is excluded from loops supporting the inner bar. Loops supporting the 
inner bar in model 02 strongly pulsate but the change in eccentricity is even more 
inconsistent among them than in the models analysed above, as shown in Figure 4. 

\subsubsection{Model 02E}
The pattern speed for the inner bar in model 02E is 70 \kmskpc. This is a new model, 
not presented in MA08, and it extends the range of models from MA08 to lower pattern 
speeds. The ring-width diagram for this model, presented in Figure 1a, looks very 
different from those for all the other models analysed in this paper. The loops, 
presented in the bottom row of Figure 2,
are also significantly different. The innermost two loops have the initial conditions 
taken from the $V$-shaped feature branching from the outer
arch in Figure 1a. These loops seem to rotate in the opposite direction to the two bars.
There is a region of small ring-widths between the left leg of the outer arch and
the inner arch. The two orbits recovered there map onto loops 3 and 4 which seem 
to be dominated by pulsation rather than rotation. The inner arch
in Figure 1a looks also different than in the models above: its left leg is considerably 
shortened and occupies the area that is chaotic in the ring-width diagrams of all the 
other models. The inner loops recovered in this area seem to be oriented perpendicularly 
to the inner bar, as if they originated from the $x_2$ orbits in the inner bar. 
None of the loops in model 02E support the inner bar and therefore they are not analyzed
further in Figures 3 and 4.

\begin{figure}[t]
\centering
\includegraphics[width=0.9\linewidth]{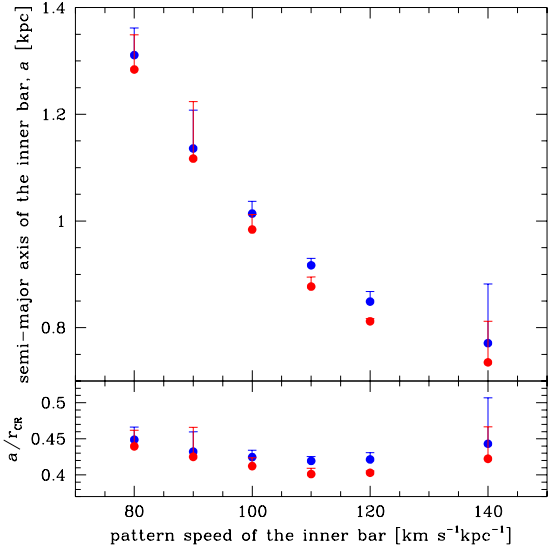}
\caption{For each of the models 02--05E, characterized by the pattern speed of
the inner bar in each model, plotted on the horizontal axis, we plot the length 
of the inner bar as determined from the extent of the loops 
that support it (upper panel), and the ratio of this length to the corotation radius
of the inner bar in each model (lower panel). Red points represent measurements
at the moment when the two bars are parallel, and blue points -- when the bars
are perpendicular. Error bars mark upper limits for the length of the bar estimated
following the method in Section 2.3. The points, from left to right, correspond to models 
02, 03, 04, 01, 05 and 05E.}
\end{figure}

\begin{figure}[h]
\includegraphics[width=0.9\linewidth]{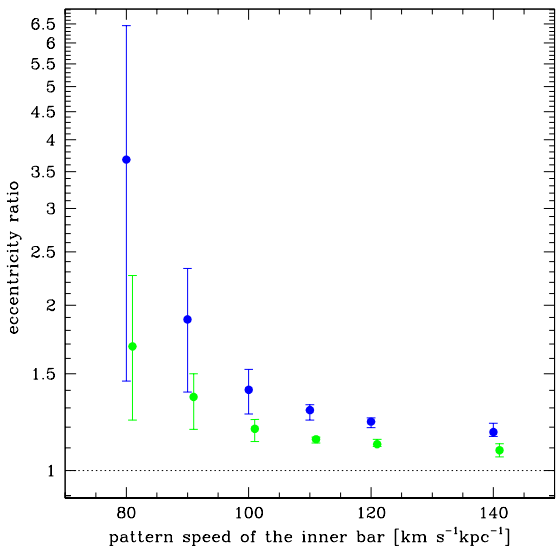}
\caption{For the loops supporting the inner bar in each of the models 02--05E (characterized 
by the pattern speed of the inner bar in each model, that is plotted on the horizontal axis) 
we plot the average ratio of the eccentricity at the moment when the two bars 
are perpendicular to the eccentricity when the two bars are parallel (blue points) and 
the average ratio of the eccentricity at the moment when the the angle 
between the two bars in the assumed potential is 45\deg\ to the eccentricity when the 
two bars are parallel (green points). Error bars mark the range of this ratio for all 
the loops that support the inner bar. The points, from left to right, correspond to 
models 02, 03, 04, 01, 05 and 05E. For clarity, data plotted in green are slightly
shifted to the right from data plotted in blue.}
\end{figure}

\subsection{Dependence of inner bar's properties on its pattern speed}
\subsubsection{The length of the inner bar}
In the fourth column of Table 1, we list the length of the semi-major axis of the inner bar, 
determined with the method described in Section 2.3 as the maximum extent of the orbits that
support it. Despite our estimation of the bar length being approximate,
there is a clear trend in the models. The extent over which the inner bar is supported by orbits 
that align with it evidently decreases with the pattern speed of this bar, as we already noticed 
when studying individual models in Section 3.1. In the upper panel of Figure 6 we plot the implied
length of
the inner bar as a function of its pattern speed, measured at the moments when the two bars are 
parallel and perpendicular to each other, as this length varies slightly when the bars rotate 
through each 
other. It is clear that the extent of orbital support for inner bars of lower pattern speed 
extends further out.

When the implied length of the semi-major axis of the inner bar is plotted in units of its 
corotation radius (Figure 6, lower panel, and Table 1, column 5), it exhibits no clear
trend with the pattern speed, and in all models the inner bar has orbital support out to
40\%--44\% of its corotation radius. The ratio of the implied bar size to its corotation radius is 
therefore remarkably constant. This result confirms the prediction of MS00, already verified by 
numerical models (e.g. Heller et al. 2007, Shen \& Debattista 2009): the inner bar in a doubly 
barred system is unlikely to extend to its corotation.

\subsubsection{The eccentricity and the pulsation of the inner bar}
In the sixth column of Table 1, we list the largest axial ratio of the loops supporting the inner 
bar at the moment when the two bars are parallel. This axial ratio is the indicator of the 
eccentricity of the bar. As we already noticed when studying individual models in Section 3.1, 
the eccentricity of the inner bar, implied by shapes of the loops that support it, decreases with
the bar's pattern speed: the faster rotating bar becomes rounder. 

In all models presented in this paper orbital support of the inner bar indicates that this
bar should pulsate as it rotates through the outer bar. The inner bar becomes rounder as it
aligns with the outer bar, while its eccentricity is largest when the two bars are perpendicular
(see Figure 3, right-hand panels), as already predicted by MS00. The pulsation of the inner bar, 
quantified by the average ratio of eccentricities of the loops at the moments when the two bars 
are parallel and perpendicular, decreases with the pattern speed of the bar (Table 1, column 7). 
Thus the lower the pattern speed of the inner bar, the more eccentric that bar is and the more it 
pulsates. In Figure 7 we plot the average ratio of eccentricities of the loops between the 
moments when the two bars are parallel and perpendicular, as well as between the moments when 
the two bars are parallel and at 45\deg. It can be seen that the average
eccentricity of the loops when the bars are perpendicular increases from its value when they
are parallel by about twice as much as it increases when the angle between the two bars is 45\deg.
It is worth pointing out that the inner bar pulsates even when it is very round: the largest
axial ratio of the loops supporting
the inner bar in model 05E is only 1.5 when the bars are parallel, but it still increases by
18\% to 1.78 when the bars are perpendicular.

For the models with the highest inner bar pattern speeds (models 01, 05 and 05E), the loops 
pulsate only slightly and in a coherent way (see Figure 4). For models with lower pattern speeds
of the inner bar, and model 02 in particular, the average pulsation is more extreme and there 
is more variation in the pulsation of individual loops (Figure 7). Thus a slowly rotating inner bar 
does not pulsate coherently: when the eccentricity of its outermost loops increases by a mere 50\%
the eccentricity of its inner loops can increase by factors larger than six.

\subsubsection{The non-uniform rotation of the inner bar}
As predicted from the orbital structure of double bars (MS00) and confirmed by numerical
models (Debattista \& Shen 2007), the inner bar in a doubly barred galaxy does not rotate 
uniformly, but its angular velocity is highest when the bars are aligned and lowest when 
the bars are perpendicular. The same is implied by the orbital models presented here: the loops 
supporting the inner bar align with the uniformly rotating major axis of the bar in the 
assumed gravitational potential when this bar is parallel or perpendicular to the outer bar, 
but they lead this axis when the inner bar departs from the alignment with the outer one, and 
trail it on the way back to the alignment (see Figure 3, left-hand panels). In particular, when
in the assumed gravitational potential the angle between the major axes of the two bars is
45\deg, the PA of the major axis of the loops supporting the inner bar is offset
from this angle to a larger value. This offset is listed in the last column of Table 1. It 
indicates the
amplitude of the variation of the pattern speed of the inner bar. If for simplicity one
assumes that this variation is sinusoidal, then the actual pattern speed of the inner bar, 
$\Omega_S^v(t)$, relates to the assumed constant pattern speed, $\Omega_S$, by 
$\Omega_S^v(t) = \Omega_S (1+2\epsilon\cos[2\Omega_S t])$, where $\epsilon$ in radians is 
the PA offset listed in Table 1.

\begin{figure}[t]
\centering
\includegraphics[width=1.0\linewidth,viewport=10 250 600 600,clip]{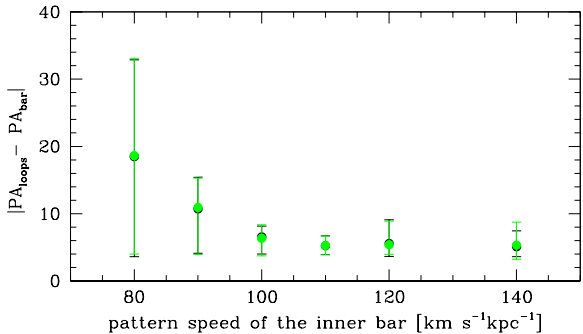}
\caption{For each of the models 02--05E (characterized by the pattern speed of the inner 
bar in each model, which is plotted on the horizontal axis) the dots mark the average offset in 
the PA of the major axis of the loops that support the inner bar from the PA
of the uniformly rotating major axis of the inner bar in the imposed gravitational
potential at the moments when the angle between the two bars in the imposed potential is 
45\deg (green) and 135\deg (black). The maximum and minimum of this offset is marked by 
the error bars. The points, from left to right, correspond to models 02, 03, 04, 01, 
05 and 05E.}
\end{figure}

In Figure 8, we plot the PA offset when the angle between the inner and the outer
bar is 45\deg\ and 135\deg. In addition to the average PA offset, we plot its 
maximum and minimum among the loops supporting the inner bar in each model, represented by 
error bars. The offset for 45\deg\ and 135\deg\ looks the same, because time is reversible in the 
equation of motion. Small range between the minimum and maximum offset indicates that the
loops remain aligned, and therefore the non-uniform rotation of the bar remains coherent. 
Thus fast inner bars (models 04, 01, 05 and 05E) rotate coherently. 
When the angle between the bars in the assumed potential is 45\deg, they lead the inner 
bar in the assumed potential by 5\deg--7\deg, which implies that the pattern speed of the 
inner bar varies by 18\%--24\% around its nominal value.

Also in Figure 8 it can be seen that slowly rotating  inner bars (models 02 and 03)
have the PA offset on average larger, and the range between its minimum and
maximum increases as the pattern speed of the inner bar is lowered. This means that the
slower the inner bar rotates, the less uniform and less coherent its rotation is.
On average, the pattern speed of the inner bar is predicted to vary around its nominal value 
by 38\% in model 03 and by 65\% in
model 02. However, as can be seen in the left-hand panels of Figure 3, in both models the
PA offset increases towards the centre, hence the rotation is least uniform
in the central parts of the inner bar. Inner loops in model 02 have the PA offset 
above 25\deg, which means that their pattern speed varies by a factor of a 
few (the no longer adequate sinusoidal approximation gives the amplitude of 90\%). 

On the other hand, rotation in the outer parts of the slowly rotating inner bars is as 
uniform as in the fast bars. In Figure 9, we show the PA offset (averaged between 
the moments when the angle between the bars in the assumed gravitational potential is 
45\deg\ and 135\deg)
for individual loops in all the models. Its value at the ends of the inner bar in models 02
and 03 (1.3 and 1.12 kpc, respectively) is as low as the lowest value for models 05 and 05E
-- just about 4\deg. It corresponds to the variation of inner bar's pattern speed by just
14\%. Interestingly, in Figure 9 can also be seen that the minimal PA offset in
each model is bound within a very small range: 3.5\deg--4.2\deg, indicated by the two
dashed lines in that figure.

\begin{figure}[t]
\centering
\includegraphics[width=1.0\linewidth,viewport=10 250 600 600,clip]{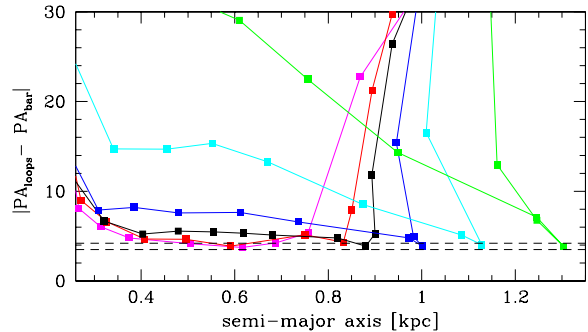}
\caption{The PA offset, averaged between the moments when the angle between
the bars in the assumed gravitational potential is 45\deg\ and 135\deg, plotted for the loops in
model 05E (pink), 05 (red), 01 (black), 04 (blue), 03 (cyan) and 02 (green), against 
their semi-major axes measured at those moments. Each loop is marked by a filled
square, with lines that connect them reflecting the sequence along the arch in Figure 1.
In each model, the minimal PA offset falls between the two black dashed 
lines drawn for 3.5\deg\ and 4.2\deg.}
\end{figure}

Given the analysis above, the rotation of the inner bar should be most coherent in models
05 and 05E. However, in Figure 8 the range of the PA offset in these models is 
larger than in model 01, whose inner bar rotates more slowly. This is because the innermost
loops that we included into the inner bar in models 05 and 05E (red and pink squares 
furthest to the left in Figure 9) have no counterpart in model 01, while being most likely 
perturbed by the inner ILR of the outer bar. If we exclude these loops, the range of
the PA offset for all the loops supporting the inner bar in models 01, 05 and
05E is just 3\deg\ in each case, hence the inner bar in these three models rotates very 
coherently.

\section{Discussion}
\subsection{The validity of orbital structure studies}
The major drawback of orbital studies is that they impose the potential without checking
whether the stars put on orbits in this potential can reproduce the imposed potential.
However, orbital studies form only a first step in dynamical analysis, by providing
a framework to fully self-consistent models, like in N-body simulations. Their role is
to identify orbits that are the backbone of the various galactic structures, conditions 
for the existence of these orbits, and to estimate the extent of these orbits as a 
function of parameters of the potential. These studies allow a quick exploration 
of the parameters of the system analyzed, as these parameters are part of the imposed
potential.

The fruitfulness of orbital studies can be seen on the example of our understanding of
the structure of classical single bars. As stated by Sellwood \& Wilkinson (1993), by 
the time of that review, only one self-consistent solution for a rapidly rotating 
two-dimensional bar existed to their knowledge (see section 4.9.1 there). Thus our insight into
bar dynamics came mostly from orbital studies, which lack self-consistency: the finding of
Contopoulos (1980) that bars should end within their corotation was confirmed by detailed studies 
of Contopoulos \& Papayannopoulos (1980), which assume the bar potential that actually extends
beyond the corotation. Athanassoula et al.(1983) related inner ring structures 
observed in some galaxies to stable circular periodic orbits near the corotation, using 
bars extending to twice their corotation radius in order to enhance the orbital response 
of interest. The orbital theory provided explanation for the existence and location of
the lobes or 'ears' at the ends of the bar (Teuben \& Sanders 1985), and the loops of the 
near-IR isophotes (Patsis et al. 1997) without modeling these structures in a 
self-consistent way. Pfenniger (1984) explained the rectangular aspect of the bulge observed 
in edge-on galaxies by the 4/1 vertical resonance, and Bureau \& Athanassoula (1999)
proposed diagnostics to determine the viewing angle with respect to the bar in an edge-on 
disk based on signatures of orbit families in the position-velocity diagrams. They were all
using models that lack self-consistency.
The method of our analysis is parallel to that of Patsis et al. (2003), who studied
the orbital response to varying pattern speed of an imposed bar of a fixed length and
axial ratio, and found that outer boxiness of the bar is favoured by fast pattern speeds, 
while a slowly rotating bar is surrounded by almost circular orbits.

The success of orbital analysis can be partially explained by the fact that the orbits
can be seen as a sum of oscillations in the gravitational potential, and therefore the 
orbital response depends more on resonances among the frequencies involved than on the 
particular shape of the gravitational potential.
Several results of orbital studies of double bars (MS00) have already been confirmed by
fully self-consistent N-body simulations (Debattista \& Shen 2007; Shen \& Debattista 2009).
Due to the presence of an outer bar, the inner bar pulsates (being thinnest when the bars 
are perpendicular and thickest when parallel), accelerates (spending more time nearly 
perpendicular to the outer bar than nearly parallel to it), and ends well within its own 
corotation. These results come from orbital studies of double barred galaxies, in which
a potential of two rigidly rotating bars is imposed. Future work, in which imposed 
potentials do not require the two bars to rotate rigidly, should bring such orbital 
studies closer to self-consistency.

\subsection{Is orbital support better in fast or slow inner bars?}
Orbital models analyzed in this paper show that the lower the pattern speed of the inner 
bar, the larger the radial extent of the orbits that support it. However, at lower pattern
speeds these orbits trap around themselves regular trajectories less efficiently than
at higher pattern speeds. It is especially true for orbits supporting outer parts 
of the inner bars, and the extreme example is shown in Figure 5. This means that although 
at low pattern speeds the bar can be longer, it can drag with itself only a small fraction
of stars, as it does not trap stellar trajectories well. On the other hand, increasing 
the pattern speed of the inner bar leads to more efficient trapping of trajectories,
but at high pattern speeds, orbits support only short and round inner bars. Thus there 
is a trade-off between the length of the inner bar and the support of this bar by regular 
trajectories. Both extremes, at the lowest and highest pattern speed of the inner bar 
are possible, as well as the intermediate solutions between them. However, there is no
preferred pattern speed for which the orbital support is 'best' in any sense. 
In particular, we confirm the finding of MA08 that the inner bar 
in resonant coupling with the outer bar is not preferred in any way.

In the high pattern speed limit (model 05E), the orbital structure implies inner bar that is short, 
round and can trap a good fraction of stars. As pointed out in Section 3.1.3, its end is not well 
defined, and the transition region to the outer bar may look like an isophotal twist
(Figure 2, top panels). The size of this transition zone increases with inner bar's
pattern speed (see left-hand columns of Figure 3 for models 05 and 05E), and one may 
expect that at still higher pattern speeds the morphology in the inner bar region is
dominated by the isophotal twist.

In the low pattern speed limit (model 02), the orbital structure implies inner bar that is long, 
eccentric and can trap only a small fraction of stars. Such bar has a well defined end in terms of 
orbits that
support it, but as these orbits do not trap stars well, the bar's end may not be defined 
that well in stellar distribution. Further decreasing of pattern speed of the inner bar 
erases the supporting orbits abruptly and completely. Thus there is a well defined lowest 
possible, critical pattern speed of the inner bar. For pattern speeds lower than that, 
there are still regular orbits within the inner bar, but their appearance is dramatically
different, and they no longer support the shape of the inner bar. Most of them remain
perpendicular to the inner bar. If the presence of such orbits were a mere consequence 
of this bar having an ILR, they should be also present in models 04, 03 and 02, as all 
these models have an ILR (see fig.9 in MA08). We searched for such 
orbits in those models, but we did not find any.  Also, if the structure of the inner 
and the outer bar were similar, families of orbits both parallel and perpendicular to the 
bar should coexist within the same model, which is not the case in our models. The abrupt
disappearance, at low pattern speeds, of orbits aligned with the inner bar, and their
replacement throughout the extent of the bar by orbits perpendicular to it, which were
absent at higher pattern speeds, reflects the nonlinear interaction between the bars 
in doubly barred galaxies.
The value of the critical pattern speed found in this paper is specific to the set of models
considered here, and it is not clear by what physical mechanism it is determined. It is close
to twice the pattern speed of the outer bar, and one can hypothesize that the low-order
resonance between the bars destroys the support of the inner bar. However, no transition in
orbital structure is seen when the pattern speed of the inner bar is three times that of the
outer bar.

\subsection{Possible dynamical interpretation of the observed morphologies}
Recently a method has been proposed for extracting inner bars from observations of 
early-type spirals by modelling and subtracting the disk, bulge and the outer bar. The
results for two galaxies have already been presented (Erwin 2010). The two extracted 
inner bars are quite different from each other. One (in NGC 1543) contains a small fraction 
of the total stellar light (4\%), its eccentricity is high (axial ratio of 4), and its
extent rather well defined, with isophotes aligned throughout the bar.
The other bar (in NGC 2859), to 
the contrary, contains a much larger fraction of the total light (10\%), its eccentricity
is lower (axial ratio of 2), and its outer part smoothly turns into a region of 
twisted isophotes outside it, which at larger radii become round and similar to those of the
subtracted components of the galaxy in the same region.

Based on our knowledge of single bars, one should not necessarily expect correlations between the 
three characteristics above (eccentricity, mass, and outer isophotal shape). It is therefore
significant that in the observations they correlate in the same way as in our two models of
extreme pattern speeds. Orbits supporting the slowly rotating inner bar in our models map onto 
loops that are eccentric, hence large axial ratio of such a bar, but they do not trap stars well,
hence small mass fraction in the bar. We also showed that because of its orbital structure the 
extent of a slow bar is well defined. The inner bar in NGC 1543 is eccentric, contains small mass 
fraction and has a well defined extent, which suggests that it rotates slowly in the sense that 
its pattern speed is close to the lowest possible one.

On the other hand, in our models orbits supporting the inner bar of high pattern speed map onto 
loops that become 
increasingly round as the pattern speed of the bar increases, therefore the observed axial 
ratio of that bar should be small. These orbits trap stars very efficiently, hence the majority 
of the stars within the extent of the bar follow that bar and the mass fraction in the bar is large. 
Outer 
loops of fast rotating inner bars become monotonically rounder as their major axes increase,
and enter a transition region that looks like an isophotal twist, so that the end of the bar
is not well defined. The inner bar of NGC 2859 is less eccentric than in NGC 1543, contains a high
mass fraction and is surrounded by an isophotal twist which suggests that the angular velocity of the
inner bar in NGC 2859 is considerably larger than its lowest dynamically possible value, and
that the structure of this inner bar is close to the structure in models of fast inner bars
presented here.

Thus, we propose that from the morphology of extracted inner bars we can discriminate
between slow bars, whose pattern speed is close to the minimal dynamically possible value, 
and fast bars, whose pattern speed is considerably above this limit (possibly two times higher).
This distinction can be based on
the characteristics listed in Table 2. Note that by slow and fast inner bars we do not
mean the ratio of the length of the bar to its corotation radius, which in our models is 
virtually constant. Our slow bars have pattern speeds in a range whose lower limit is set
by no orbital support for the inner bar, while our fast bars become increasingly rounder as
their angular velocity increases, and eventually they cannot be distinguished from
axisymmetric components of the galaxy.

\begin{table}[t]
\begin{center}
\caption{Morphological characteristics of slow and fast inner bars in doubly barred galaxies}
\vspace{5mm}
\begin{tabular}{lll}
\tableline\tableline
             & slow inner bar       & fast inner bar\\
\tableline
mass fraction& small                & large \\
eccentricity & large                & small \\
end of bar   & well defined by      & unclear, turns into\\
             & aligned isophotes    & twisted isophotes\\
\tableline
\end{tabular}
\end{center}
\end{table}
\section{Conclusions}

In this paper, we studied the orbital support of the inner bar in seven models of 
double-barred galaxies, with the angular velocity of the inner bar different in each 
model. The aim of this work was to analyze how this support changes with the bar's 
pattern speed. 

The new result of this analysis is that the pattern speed of the 
inner bar cannot be arbitrarily low. When this pattern speed drops below a certain
threshold, the family of loops that support the inner bar 
is completely wiped out. It is instantly replaced by a family of loops perpendicular 
to that bar, possibly related to the $x_2$ orbits in the {\it inner} bar. We find no 
models in which both families of loops coexist. In addition to the lower limit 
for the angular velocity of the inner bar, set by an abrupt destruction of orbits 
that support it, there is a soft upper limit, which comes from the bar becoming 
increasingly rounder as its angular velocity increases, so that it no longer is
a distinct dynamical feature. These limits apply only to double bars
that rotate in the same direction, as only such bars were considered in this paper. 
Maciejewski (2008) showed that in counter-rotating double bars, inner bars are 
supported by loops corresponding to a different orbital family ($x_4$), and their 
dynamically possible angular velocities may not be limited in a similar way.

The models presented in this paper show that the pattern speed of the inner bar
has a significant impact on its structure and dynamics. Earlier studies were
limited to single models of double bars, and this work extends what we already know 
based on those studies. In particular, the predictions of orbital models by MS00, 
verified by the $N$-body simulations of Debattista \& Shen (2007) and Shen \& Debattista 
(2009), are extended as follows:
\begin{enumerate}
\item Confirming that the inner bar should end well within its corotation radius,
we show that the orbital support of the inner bar extends further out in radius for
lower pattern speeds than it does for higher ones. 
However, lower pattern speed means larger corotation
radius, and the ratio of the extent of orbital support of the inner bar to
its corotation radius remains remarkably constant for the models constructed
here. For low pattern speeds of the inner bar, double-frequency orbits supporting 
outer parts of that bar trap trajectories that do not occupy large volumes 
of phase-space, hence provide only a limited support for the bar.
\item Confirming that the inner bar pulsates as it rotates through the outer bar,
we show that the lower the pattern speed of the inner bar, the more eccentric the 
loops that support it, and the more they pulsate as the bars rotate through each other.
\item Confirming that the rotation of the inner bar in a doubly barred galaxy is 
not uniform, we show that when the angular velocity of the inner bar is small, 
no more than 30\% above the threshold below which the loops that support it are wiped 
out, these loops do not rotate coherently: the angular velocity of the inner loops
varies severely, much more than that of the outer loops. On the other hand, faster 
inner bars rotate coherently: the loops that support the inner bar remain aligned
within a few degrees of each other as this bar rotates through the outer bar. The rotation
of such bars is quite uniform as well: the common angular velocity of the loops varies only 
by about 20\% around its average value. 
\end{enumerate}

We find no dynamically preferred pattern speed of the inner bar. In particular, there
is no evidence of minimizing chaos at resonant coupling between the two bars. However,
there is a trade-off
in the properties of the bar instead. At large pattern speeds, loops supporting the inner 
bar trap trajectories that occupy large volume of phase-space, so the inner bar can 
contain a large fraction of stars, but it is also short 
and round. At small pattern speeds, the inner bar is longer, more eccentric, but it pulsates 
and accelerates more, and the volume of chaotic zones increases, hence the bar cannot trap
enough stars to make its mass high. This correlation between mass, eccentricity and the
isophotal shape seems to be reflected by the recent observational decomposition of inner bars
in doubly barred galaxies.

\acknowledgments

We would like to thank Peter Erwin for commenting on this work, and
Lia Athanassoula for inspiring discussions. This work was partially supported by the Polish 
Committee for Scientific Research as a research project 1 P03D 007 26 in the years 2004--2007.
WM acknowledges Academic Fellowship EP/E500587/1 from Research Councils UK.

\clearpage

\end{document}